\documentclass[aps,preprint,nofootinbib,floatfix,a4paper]{revtex4-1}
\usepackage{graphicx}
\usepackage[latin1]{inputenc}
\usepackage{amsmath,amssymb}
\usepackage{hyperref}
\usepackage{epstopdf}
\usepackage{geometry}    
\usepackage{caption}
            
\def\be{\begin{equation}}
\def\ee{\end{equation}}
\def\bea{\begin{eqnarray}}
\def\eea{\end{eqnarray}}

\begin{document}

\title{Lepton Flavor Violating Decays of Neutral Higgses\\ 
in Extended Mirror Fermion Model}

\author{
Chia-Feng Chang $^1$,
Chia-Hung Vincent Chang $^2$, 
Chrisna Setyo Nugroho $^2$ and Tzu-Chiang Yuan $^{3,4}$
}

\affiliation{
$^1$Department of Physics, National Taiwan University, Taipei 116, Taiwan\\
$^2$Department of Physics, National Taiwan Normal University, Taipei 116, Taiwan\\
$^3$Institute of Physics, Academia Sinica, Nangang, Taipei 11529, Taiwan\\
$^4$Physics Division, National Center for Theoretical Sciences, Hsinchu, Taiwan
}

\date{\today}                                          

\begin{abstract}

We perform the one-loop induced charged lepton flavor violating decays of the neutral Higgses
in an extended mirror fermion model with non-sterile electroweak-scale right-handed neutrinos
and a horizontal $A_4$ symmetry in the lepton sector.
We demonstrate that for the 125 GeV scalar $h$ there is tension between the recent LHC result  
${\cal B}(h \to \tau \mu) \sim $ 1\% and 
the stringent limits on the rare processes $\mu \to e \gamma$ and $\tau \to (\mu$ or $e) \gamma$ 
from low energy experiments.

\end{abstract}

\maketitle

\section{Motivation}

As is well known, lepton and baryon number are accidental global symmetries in the fundamental Lagrangian 
of Standard Model (SM). Processes like $\mu \to e \gamma$, $p \to e \gamma$, {\it etc}
that violating either one (or both) of these two quantum numbers are thus strictly forbidden
in the perturbation calculations of SM. Experimental limits for these processes are indeed very stringent. 
For example, from Particle Data Group \cite{PDG2015},
we have the following bounds
\begin{equation}
\label{MEG}
{\cal B}(\mu^- \to e^- \gamma) < 5.7 \times 10^{-13} \; (90 \, \% \, {\rm CL}) \; , 
\end{equation}
and
\begin{equation}
\label{protondecay}
\tau (p \to e^+ \gamma) > 670 \times 10^{30} \; {\rm years} \; . 
\end{equation}

Search for  lepton flavor violating (LFV) Higgs decay
$h \to \tau \mu$ at hadron colliders was proposed some time ago \cite{Han:2000jz}.
Recently both ATLAS~\cite{Aad:2015gha} and CMS ~\cite{Khachatryan:2015kon} 
experiments at the Large Hadron Collider (LHC) have reported
the following best fit branching ratios
\begin{equation}
\label{LHCResults}
{\cal B}(h \to \tau \mu) = \left\{
\begin{array}{l}
0.84^{+0.39}_{-0.37} \, \% \; (2.4 \sigma) \;   [{\rm CMS}] \; , \\
0.77 \pm 0.62 \, \% \; (1.2 \sigma) \;   [{\rm ATLAS}] \; .
\end{array}
\right.
\end{equation}
However, at 95\% confidence level (CL), the following upper limits can be deduced 
\begin{equation}
\label{LHCLimits}
{\cal B}(h \to \tau \mu) = \left\{
\begin{array}{l}
 < 1.85 \, \%  \; (95 \, \% \, {\rm CL}) \; [{\rm ATLAS}] \; ,\\
 < 1.51 \, \%  \; (95 \,\% \, {\rm CL}) \; [{\rm CMS}] \; .
\end{array}
\right.
\end{equation}
Despite low statistical significance the above best fit results in Eq.~(\ref{LHCResults}) are somewhat surprising since 
for a 125 GeV Higgs the branching ratio for this mode is about $3.6 \times 10^{-6}$ in the SM augmented by
the minuscule neutrino mass terms.
A positive measurement of this branching ratio in the near future at the percent level 
would be a clear indication of new physics beyond the SM.

On the other hand, we have stringent limits for LFV radiative decays 
like $\mu \to e \gamma$ in Eq.~(\ref{MEG}) as well as 
\begin{eqnarray}
\label{BaBarResults}
{\cal B}(\tau \to \mu \gamma) & < & 4.4 \times 10^{-8}  \; , \\
{\cal B}(\tau \to e \gamma) & < & 3.3 \times 10^{-8}   \; , 
\end{eqnarray}
both at 90\% CL from the low energy data of BaBar experiment \cite{Aubert:2009ag}.

Over the years, many authors had studied the flavor changing neutral current Higgs decays 
$h \to {\overline f_i} f_j$ in both the SM and its various extensions. 
For a recent updated calculation on $h \to {\overline q_i} q_j$ in the SM we refer the readers to 
\cite{Benitez-Guzman:2015ana} and references therein. 
For earlier calculations for the leptonic case with large Majorana neutrino masses, 
see for example~\cite{Pilaftsis:1992st,Korner:1992zk}.
Recently large flux of works on new physics implications for the LHC result Eq.~(\ref{LHCResults})
is easily noticed 
\cite{Dorsner:2015mja,
Han:2016bvl,
Sher:2016rhh,
Buschmann:2016uzg,
Crivellin:2016vjc,
Bizot:2015qqo,
Hue:2015fbb,
Cline:2015lqp,
Zhang:2015csm,
Omura:2015xcg,
Benbrik:2015evd,
Aloni:2015wvn,
Arganda:2015uca,
Cai:2015poa,
Baek:2015fma,
Chen:2015nta,
Baek:2015mea,
Kosnik:2015lka,
Liu:2015oaa,
Botella:2015hoa,
Arganda:2015naa,
Mao:2015hwa,
Altunkaynak:2015twa,
He:2015rqa,
Chiang:2015cba,
Altmannshofer:2015esa,
Cheung:2015yga,
Das:2015zwa,
Heeck:2014qea,
Sierra:2014nqa,
Harnik:2012pb}.

In \cite{Hung:2015hra}, an up-to-date analysis of a previous calculation \cite{Hung:2007ez} 
of $\mu \to e \gamma$  in a class of mirror fermion models with 
non-sterile electroweak scale right-handed neutrinos \cite{Hung:2006ap} was presented
for an extension of the models with a horizontal $A_4$ symmetry in the lepton sector \cite{Hung:2015nva}.
It was demonstrated in \cite{Hung:2015hra} that although there exists parameter space relevant to electroweak physics
to accommodate the muon magnetic dipole moment anomaly 
$\Delta a_\mu = 288(63)(49)\times 10^{-11}$ \cite{PDG2015}, 
the current low energy limit  Eq.~(\ref{MEG}) on the branching ratio 
${\cal B}(\mu \to e \gamma)$ from MEG experiment  
\cite{Adam:2013mnn}  has disfavored those regions of parameter space.
 
In this work, we present the calculation of LFV decay of the neutral Higgses 
in an extended mirror fermion model.
In Section 2, we briefly review the extended model and 
show the relevant interactions that may lead to the LFV decays of the neutral Higgses in the model.
In Section 3, we present our calculation. 
Numerical results are given in Section 4.
We conclude in Section 5.
Detailed formulas for the loop amplitudes are given in the Appendix.

\section{The Model and its relevant interactions}

\begin{figure}[hptb!]
\includegraphics[width=0.45\textwidth]{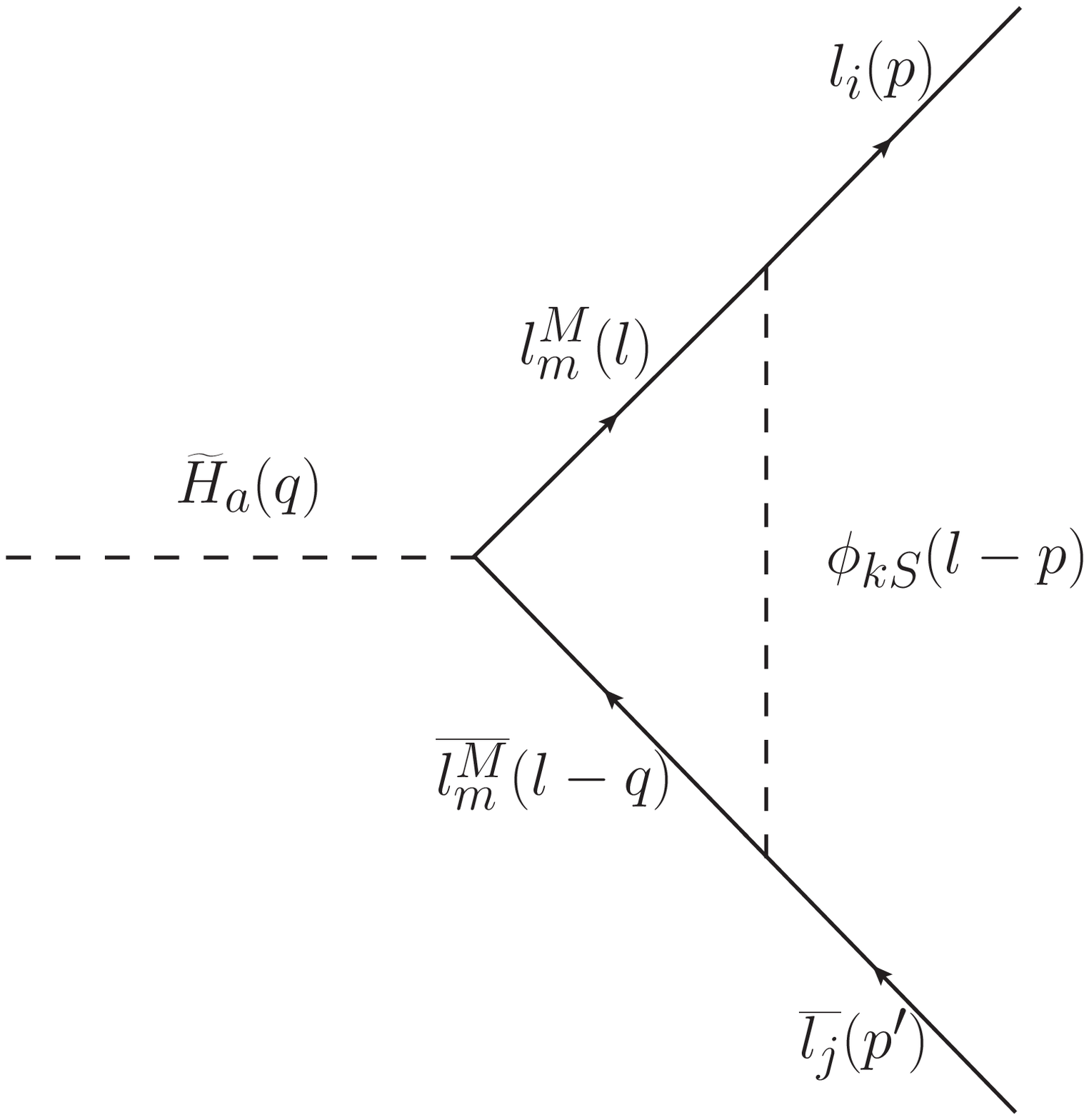} 
\caption{One-loop induced Feynman diagram for ${\widetilde H}_a(q) \to  l_i(p) + {\overline l_j} (p')$ 
in EW-scale $\nu_R$ model. The other two 1-particle reducible diagrams  
corresponding to the wave function renormalization of the external fermion lines are not shown.}
\label{FeynDiag}
\end{figure}

In the original mirror fermion model \cite{Hung:2006ap}, while the gauge group is the same as SM, every left-handed 
(right-handed) SM fermion has a right-handed (left-handed) mirror partner, and the scalar sector consists of one SM Higgs doublet $\Phi$, one singlet $\phi_{0S}$ and two triplets $\xi$ and $\tilde \chi$  $\acute{a}$ 
{\it la} Georgi-Machacek \cite{Georgi:1985nv,Chanowitz:1985ug}. 
One peculiar feature of the model  
is that the right-handed neutrinos are non-sterile. They are paired up with
right-handed mirror charged leptons to form electroweak doublets. This arrangement allows for the electroweak seesaw mechanism~\cite{Hung:2006ap}: a small vacuum expectation value (VEV) of 
the scalar singlet $\phi_{0S}$ provides Dirac masses for the light neutrinos, while a VEV with electroweak size 
of the Georgi-Machacek triplets provide Majorana masses for the right-handed neutrinos.

Recently, the original model \cite{Hung:2006ap} is augmented with an additional mirror Higgs doublet $\Phi_M$ 
in \cite{Hoang:2014pda} so as to accommodate the 125 GeV Higgs observed at the LHC.
In additional to the original singlet scalar $\phi_{0S}$,
a $A_4$ triplet of scalars $\{ \phi_{kS} \}$ $ (k=1,2,3)$ is introduced in \cite{Hung:2015nva} 
to implement a horizontal $A_4$ symmetry in the lepton sector which may lead to interesting 
lepton mixing effects. The three generations of SM leptons are assigned to be in a triplet of $A_4$ 
while the SM Higgs doublet and the triplets are singlets of $A_4$.

We will consider both extensions with $A_4$ symmetry~\cite{Hung:2015nva} and 
mirror Higgs doublet~\cite{Hoang:2014pda} in our calculation. 
The relevant Feynman diagram for LFV Higgs decay in the extended mirror model 
is one-loop induced  and is shown in Fig.~(\ref{FeynDiag}). 
The relevant interactions are all of Yukawa couplings. The first one is for the singlet $\phi_{0S}$ and triplet $\phi_{kS} (k=1,2,3)$~\cite{Hung:2015hra}
\bea
{\mathcal L}_{S} & = & - \sum_{k=0}^3 \sum_{i,m=1}^3  \left( \bar{l}_{Li} \, {\cal U}^{L \, k}_{im} l^M_{R m}  
+ \bar{ l}_{R i}  \, {\cal U}^{R \, k}_{im} l^M_{Lm} \right) \phi_{kS} + {\rm H.c.} 
\eea
where $l_{Li}$ and $l_{Ri}$ are SM leptons, $l^M_{Rm}$ and $l^M_{Lm}$ are mirror leptons 
($i,m$ are generation indices); ${\cal U}^{L \, k}_{im}$ 
and ${\cal U}^{R \, k}_{im}$ are the coupling coefficients given by
\bea
{\cal U}^{L \, k}_{im} 
& \equiv &  \left( U^\dagger_{\rm PMNS}  \cdot   M^k \cdot  U^{l^M}_{\rm PMNS} \right)_{im} \;\; , \nonumber \\
& = & \sum_{j,n = 1}^3 \left( U^\dagger_{\rm PMNS} \right)_{i j}   M^k_{jn}  
\left(  U^{M}_{\rm PMNS} \right)_{nm} \; \; , \\
{\cal U}^{R \, k}_{im} 
& \equiv & \left( U^{\prime \, \dagger}_{\rm PMNS} \cdot   M^{\prime \, k} \cdot U^{\prime \, l^M}_{\rm PMNS} \right)_{im} \; \; , \nonumber \\
& = & \sum_{j,n = 1}^3 \left( U^{\prime \, \dagger}_{\rm PMNS} \right)_{i j}   M^{\prime \, k}_{jn}  
\left(  U^{\prime \, M}_{\rm PMNS} \right)_{nm} \; \; ,
\eea
where the matrix elements for the four matrices $M^k (k=0,1,2,3)$ are listed in Table~I and
$M^{\prime \, k}_{jn}$ can be obtained from $M^{k}_{jn}$ with the following substitutions
for the Yukawa couplings $g_{0S} \to g^\prime_{0S}$ and  $g_{1S} \to g^\prime_{1S}$~\cite{Hung:2015hra};
$U_{\rm PMNS}$ is the usual neutrino mixing matrix defined as
\be
\label{PMNS}
U_{\rm PMNS}=  U_{\nu}^{\dagger} U^{l}_{L} \, ,
\ee
and its mirror and right-handed counter-parts  $U^{M}_{\rm PMNS} $, $U^\prime_{\rm PMNS}$ 
and $U^{\prime M}_{\rm PMNS} $  are defined analogously as
\be
\label{MPMNS}
U^{M}_{\rm PMNS} = U^{\dagger}_{\nu} U^{l^M}_R \, ,
\ee
\be
\label{UPMNSprime}
U^\prime_{\rm PMNS}=  U_{\nu}^{\dagger} U^{l}_{R} \, ,
\ee
and
\be
\label{UPMNSMirrorprime}
U^{\prime M}_{\rm PMNS} = U^{\dagger}_{\nu} U^{l^M}_L \,,
\ee
where $U^{l}_{R}$ and $U^{l^M}_L$ are the unitary matrices relating the gauge eigenstates 
(fields with superscripts 0) and the mass eigenstates
\be
\label{eigenstate}
l^{0}_{L,R}= U^{l}_{L,R} l_{L,R} \;\; , \;\;\; \;\;  l^{M,0}_{R,L} =  U^{l^M}_{R,L} l^{M}_{R,L} \; ,
\ee
and
\be
\label{UCW}
U_\nu = U^\nu_L = U^\nu_R = 
\frac{1}{\sqrt{3}}
\left(
  \begin{array}{cccc}
  1 & 1 & 1 \\
  1 & \omega^2 & \omega \\
  1 & \omega & \omega^2 \\
  \end{array}
\right) \; ,
\ee
where $\omega \equiv \exp (i 2\pi /3)$ entered in the multiplication rules of $A_4$. 
The matrix in Eq.~(\ref{UCW}) was first discussed by Cabibbo and also by Wolfenstein in the context
of CP violation in three generations of neutrino oscillations \cite{CW}.

\begin{table}
\label{M}
\caption{Matrix elements for $M^k (k=0,1,2,3)$ where $\omega \equiv \exp(i 2\pi /3)$ and $g_{0S}$ and $g_{1S}$ are Yukawa couplings.}
\begin{tabular}{|c|c|}
\hline
$M_{jn}^k$ & Value \\
\hline
\hline
$M^0_{12}, M^0_{13}, M^0_{21}, M^0_{23}, M^0_{31}, M^0_{32}$ & 0 \\
$M^0_{11},  M^0_{22}, M^0_{33}$ & $g_{0S}$ \\
$M^1_{11},  M^2_{11}, M^3_{11}$  & $\frac{2}{3} \mathrm{Re} \left( g_{1S} \right)$ \\ 
$M^1_{22}, M^2_{22}, M^3_{22}$  &  $\frac{2}{3} \mathrm{Re} \left( \omega^* g_{1S} \right)$  \\
$M^1_{33}, M^2_{33}, M^3_{33}$  & $\frac{2}{3} \mathrm{Re} \left( \omega g_{1S} \right)$  \\ 
$M^1_{12}, M^1_{21}$ & $\frac{2}{3} \mathrm{Re} \left( \omega g_{1S} \right)$ \\
$M^2_{12}, M^3_{21}$ & $\frac{1}{3} \left( g_{1S} + \omega g^*_{1S} \right)$ \\
$M^3_{12}, M^2_{21}$ & $\frac{1}{3} \left( g^*_{1S} + \omega^* g_{1S} \right)$ \\ 
$M^1_{13},  M^1_{31}$ & $\frac{2}{3} \mathrm{Re} \left( \omega^* g_{1S} \right)$  \\\
$M^2_{13},  M^3_{31}$ & $\frac{1}{3} \left( g_{1S} + \omega^* g^*_{1S} \right)$ \\
$M^3_{13},  M^2_{31}$ & $\frac{1}{3} \left( g^*_{1S} + \omega g_{1S} \right)$ \\
$M^1_{23}, M^1_{32}$ & $\frac{2}{3} \mathrm{Re} \left( g_{1S} \right)$ \\
$M^2_{23}, M^3_{32}$ & $\frac{2 \omega^*}{3} \mathrm{Re} \left( g_{1S} \right)$ \\
$M^3_{23}, M^2_{32}$ & $\frac{2 \omega}{3} \mathrm{Re} \left( g_{1S} \right)$  \\
\hline
\end{tabular}
\end{table}

The second Yukawa interaction is for the couplings of neutral Higgses 
with the SM fermion pairs and the mirror fermion pairs. 
It was shown in \cite{Hoang:2014pda} that the physical neutral Higgs states 
$({\widetilde H}_1 , {\widetilde H}_2, {\widetilde H}_3 )$~\footnote{We note that
$({\widetilde H}_1 , {\widetilde H}_2, {\widetilde H}_3 )$ was denoted as 
$({\widetilde H} , {\widetilde H}^\prime, {\widetilde H}^{\prime\prime} )$ respectively in \cite{Hoang:2014pda}.} 
are in general mixture of the
unphysical neutral Higgs states $(H^0_1 , H^0_{1M} , H^{0\prime}_1)$
via an orthogonal transformation $O$~\cite{Hoang:2014pda}:
\begin{eqnarray}
\label{O}
\begin{pmatrix}
{\widetilde H}_1\\
{\widetilde H}_2\\
{\widetilde H}_3
\end{pmatrix}
& = & 
\begin{pmatrix}
a_{1,1} & a_{1,1M} & a_{1,1^\prime}\\
a_{1M,1} & a_{1M,1M} & a_{1M,1^\prime}\\
a_{1^\prime,1} & a_{1^\prime,1M} & a_{1^\prime,1^\prime}
\end{pmatrix}
\cdot
\begin{pmatrix}
H^0_1 \\
H^0_{1M} \\ 
H^{0\prime}_1
\end{pmatrix} \nonumber \\
& \equiv & O \cdot \begin{pmatrix}
H^0_1 \\
H^0_{1M} \\ 
H^{0\prime}_1
\end{pmatrix}
\;\; ,
\end{eqnarray}
where $H^0_1$ and $H^0_{1M}$ are the neutral components of the SM 
Higgs and mirror Higgs doublets 
respectively, and $H^{0\prime}_1$ is linear combination of the neutral components in the 
Georgi-Machacek triplets. The couplings of the physical Higgs 
${\widetilde H}_a$ with a pair of SM fermions $f$ and a pair of mirror fermions $f^M$ 
are given by~\cite{Hoang:2014pda} 
\begin{equation}
\label{LHiggses}
{\cal L}_{\widetilde H} = - \frac{g}{2 m_W} 
\sum_{a,f} {\widetilde H}_a  \left\{ m_f \frac{O_{a1}}{s_2} {\overline f} f 
+ m_{f^M} \frac{O_{a2}}{s_{2M}}  \overline{f^M} f^M
\right\} \; ,
\end{equation}
where $g$ is the $SU(2)_L$ weak coupling constant; $m_W$ is the $W$ boson mass; 
$O_{a1}$ and $O_{a2}$ are the first and second columns of the above 
orthogonal matrix $O$ in Eq.~(\ref{O}); $s_2$, $s_{2M}$ and $s_M$ are mixing angles defined by
\begin{eqnarray}
s_2 & =& \frac{v_2}{v} \; ,\\
s_{2M} & = & \frac{v_{2M}}{v} \; , \\
s_M & = & \frac{2 \sqrt 2 v_M}{v} \; ,
\end{eqnarray}
with $v=\sqrt{v_2^2 + v_{2M}^2 + 8 v_M^2}=246$ GeV, 
where $v_2$, $v_{2M}$ and $v_M$ are the VEVs of the Higgs doublet, 
mirror Higgs doublet and Georgi-Machacek triplets respectively.
For the original mirror model \cite{Hung:2006ap}, one can simply set
${\widetilde H}_1 \to H^0_1\equiv h$, $O_{11}/s_2$ and $O_{12}/s_{2M} \to 1$,  
and drop all other terms with $a \neq 1$ in Eq.~(\ref{LHiggses}).

\section{The Calculation}


The matrix element for the process ${\widetilde H}_a(q) \to  l_i(p) + {\overline l_j} (p')$ (Fig.~\ref{FeynDiag})
can be written as 
\begin{equation}
\label{AmpCD}
i {\cal M}  =  i \frac{1}{16 \pi^2}  
\overline{u_i}(p) \left( C^{aij}_L P_L + C^{aij}_R P_R \right) v_j (p') \; , 
\end{equation}
where $P_{L,R}=(1\mp \gamma_5)/2$ are the chiral projection operators. In terms of scalar and pseudoscalar couplings 
the above amplitude can be rewritten as
\begin{equation}
\label{AmpAB}
i {\cal M}  =  i \frac{1}{16 \pi^2} \overline{u_i}(p) \left( A^{aij} + i B^{aij} \gamma_5 \right) v_j (p') \; ,
\end{equation}
where
\begin{equation}
\label{AB2C}
A^{aij} = \frac{1}{2} \left (C^{aij}_L + C^{aij}_R \right) \;\; , \; \; B^{aij} = \frac{1}{2i} \left (C^{aij}_R - C^{aij}_L \right) \; .
\end{equation}
The partial decay width is given by
\begin{eqnarray}
\Gamma^{aij} &=& \frac{1}{2^{11} \pi^5} m_{\widetilde H_a}
\lambda^{\frac{1}{2}} \left( 1 , \frac{m_i^2}{m_{\widetilde H_a}^2}, \frac{m_j^2}{m_{\widetilde H_a}^2} \right)  \nonumber \\
&  & \times \left[ \vert A^{aij} \vert^2 \left( 1 - \frac{(m_i + m_j)^2}{m_{\widetilde H_a}^2} \right) 
+ \vert B^{aij} \vert^2  \left( 1 - \frac{(m_i - m_j)^2}{m_{\widetilde H_a}^2} \right)  
\right] \; ,
\end{eqnarray}
where $\lambda (x,y,z) = x^2 + y^2 + z^2 - 2 ( xy + yz + zx )$. 
The one-loop induced coefficients $A^{aij}$ and $B^{aij}$ are related to 
$C^{aij}_L$ and $C^{aij}_R$ according to Eq.~(\ref{AB2C}). The formulas for the
latter are given in the Appendix.

We now comment on the divergent cancellation in the calculation.
For the original mirror model \cite{Hung:2006ap} in which there is only one Higgs doublet with 
Yukawa couplings to the SM fermions and to the mirror fermions that are differ only by the corresponding
fermion masses, the divergence in the one-loop diagram in Fig.~(\ref{FeynDiag}) will cancel with those 
in the two 1-particle reducible diagrams associated with wave function renormalization.
On the other hand, for the extended model \cite{Hoang:2014pda} these divergences do not cancel each other.
Recall that in the extended model, besides the SM Higgs doublet
an additional mirror Higgs doublet was introduced. Both Higgs doublets can 
then couple to the SM fermions and may lead to LFV decay of the Higgses at tree level. 
In \cite{Hung:2006ap}, a global $U(1) \times U(1)$ symmetry was employed 
such that the SM Higgs doublet only couples to the SM fermions, 
while the mirror Higgs doublet only couples to the mirror fermions. Hence there will be no tree level 
LFV vertices for the SM Higgs decays into SM fermions. However this global symmetry is broken by a term in the
scalar potential. This term also provide the Higgs mixings in Eq.~(\ref{O}) that eventually 
responsible to LFV decays of the Higgses in the extended model. 
Due to renormalizability, the presence of this symmetry breaking term 
in the scalar potential forces one to reintroduce the Yukawa terms that are forbidden by the symmetry.  
Hence tree level LFV decays of the Higgses are generally present in the extended model. According to the general 
analysis in \cite{Harnik:2012pb} such tree level LFV couplings 
are constrained to be quite small by low energy data.
For our purpose, we will assume these tree level LFV couplings are vanishing small and the main reason for their
existence is to provide counter terms to absorb the divergences in the calculation in the extended model.
The results of $C^{aij}_{L,R}$ should then be regarded as renormalized quantities.

The amplitude for $l_i \to l_j \gamma$ in the extended model can be found in \cite{Hung:2015hra}.

\section{Numerical Analysis}

We will focus on the case of lightest neutral Higgs $\widetilde H_1 \to \tau \mu$ with $\widetilde H_1$ 
identified as the 125 GeV Higgs, and adopt the following strategy which has been used 
in \cite{Hung:2015hra} for the numerical analysis of $\mu \to e \gamma$:

\begin{itemize}
\item
Two scenarios were specified according to the following forms of the three unknown mixing matrices:\\

Scenario 1 (S1):  $U^\prime_{\rm PMNS} = U^M_{\rm PMNS} = U^{\prime M}_{\rm PMNS} = U_\nu = {\rm Eq.}~(\ref{UCW})$

Scenario 2 (S2): $U^\prime_{\rm PMNS} = U^M_{\rm PMNS} = U^{\prime M}_{\rm PMNS} = U_{\rm PMNS}$, 
where
\bea
U^{\rm NH}_{\rm PMNS} =
\left(
\begin{array}{ccc}
0.8221 & 0.5484 & -0.0518 + 0.1439 i \\
-0.3879 + 0.07915 i & 0.6432 + 0.0528 i & 0.6533 \\
0.3992 + 0.08984 i & -0.5283 + 0.05993 i & 0.7415
\end{array}
\right) \nonumber
\eea
and
\bea
U^{\rm IH}_{\rm PMNS} =
\left(
\begin{array}{ccc}
0.8218 & 0.5483 & -0.08708 + 0.1281 i \\
-0.3608 + 0.0719 i & 0.6467 + 0.04796 i & 0.6664 \\
0.4278 + 0.07869 i & -0.5254 + 0.0525 i & 0.7293
\end{array}
\right) \nonumber
\eea
for the neutrino masses with normal and inverted hierarchies respectively. The Majorana phases have been ignored 
in the analyses. 
For each scenario, we consider these two possible solutions for the $U_{\rm PMNS}$.
Due to the small differences between these two solutions, we expect our results 
are not too sensitive to the neutrino mass hierarchies.

\item
All Yukawa couplings $g_{0S}, g_{1S}, g^{\prime}_{0S}$ and $g_{1S}^{\prime}$ are assumed to be real.
For simplicity, we will assume
$g_{0S} = g^\prime_{0S}$, $g_{1S} = g^\prime_{1S}$ and study the following 6 cases:
\begin{enumerate}
\item[(a)]
$g_{0S} \neq 0, \; g_{1S} = 0$. The $A_4$ triplet terms are switched off.
\item[(b)]
$g_{1S} = 10^{-2} \times g_{0S}$. The $A_4$ triplet couplings are merely one percent of the singlet ones.
\item[(c)]
$g_{1S} = 10^{-1} \times g_{0S}$. The $A_4$ triplet couplings are 10 percent of the singlet ones.
\item[(d)]
$g_{1S} = 0.5 \times g_{0S}$. The $A_4$ triplet couplings are one half of the singlet ones.
\item[(e)]
$g_{1S} = g_{0S}$. Both $A_4$ singlet and triplet terms have the same weight. 
\item[(f)]
$g_{0S} = 0, \; g_{1S} \neq 0$. The $A_4$ singlet terms are switched off.
\end{enumerate}

\item
For the masses of the singlet scalars $\phi_{kS}$, we take
$$
m_{\phi_{0S}} : m_{\phi_{1S}} : m_{\phi_{2S}} : m_{\phi_{3S}}= M_S : 2 M_S : 3 M_S : 4  M_S
$$
with a fixed common mass $M_S = 10$ MeV. As long as $m_{\phi_{kS}} \ll m_{l^M_m}$, our results will not
be affected much by this assumption. 

\item
For the masses of the mirror lepton $l^M_m$, we take 
$$
m_{l^M_m} = M_{\rm mirror} + \delta_m
$$
with $\delta_1 = 0$, $\delta_2 = 10$ GeV, $\delta_3 = 20$ GeV and vary the
common mass $M_{\rm mirror}$. 

\item
As shown in \cite{Hoang:2014pda}, the 125 GeV scalar resonance $h$ discovered at the LHC 
identified as the lightest state ${\widetilde H}_1$ can be belonged to the 
{\it Dr. Jekyll} scenario in which the SM Higgs doublet $H_1^0$ has a major component 
or the {\it Mr. Hyde} scenario in which it is an impostor with $H_1^0$ only a sub-dominant component.
Of all the explicit examples found for both of these scenarios, 
we will study the two following cases~\cite{Hoang:2014pda}:
\begin{itemize}

\item 
{\it Dr. Jekyll} case (Eq.~(50) of  \cite{Hoang:2014pda}):

\begin{equation}
\label{DrJekyllO}
O = \begin{pmatrix}
0.998 & -0.0518 & -0.0329 \\
0.0514 & 0.999 & -0.0140 \\
0.0336 & 0.0123 & 0.999
\end{pmatrix}\; ,
\end{equation}
with  Det$(O)=+1$,   $m_{\widetilde H_1} = $ 125.7 GeV, $m_{\widetilde H_2} = $ 420 GeV, 
$m_{\widetilde H_3} = $ 601 GeV, $s_2=0.92$, $s_{2M}=0.16$ and $s_M = 0.36$.
In this case, 
\begin{equation}
\label{DrJekyllHiggses}
h \equiv {\widetilde H}_1 \sim H^0_1\, , \;\; {\widetilde H}_2 \sim H^0_{1M}\, , \;\; {\widetilde H}_3 \sim H^{0\prime}_1 \; .
\end{equation}
Hence the 125 GeV Higgs identified as ${\widetilde H}_1$ is composed mainly of 
the neutral component of the SM doublet in this scenario.

\item
{\it Mr. Hyde} case (Eq.~(55) of  \cite{Hoang:2014pda}):

\begin{equation}
\label{MrHydeO}
O = \begin{pmatrix}
0.187 & 0.115 & 0.976 \\
0.922 & 0.321 & -0.215 \\
0.338 & -0.940 & 0.046
\end{pmatrix}\; , 
\end{equation}
with  Det$(O)=-1$, $m_{{\widetilde H}_1} = $ 125.6 GeV, $m_{{\widetilde H}_2} = $ 454 GeV, $m_{\widetilde H_3} = $ 959 GeV, $s_2=0.401$, $s_{2M}=0.900$ and $s_M = 0.151$.
In this case, 
\begin{equation}
\label{MrHydeHiggses}
h \equiv {\widetilde H}_1 \sim H^{0\prime}_1\, , \;\; {\widetilde H}_2 \sim H^0_1\, , \;\; {\widetilde H}_3 \sim H^0_{1M} \; .
\end{equation}
Hence the 125 GeV Higgs identified as ${\widetilde H}_1$ is an impostor in this scenario; it is mainly 
composed of the two neutral components in the Georgi-Machacek triplets.

\end{itemize}
\end{itemize}

\begin{figure}[h!]
\begin{minipage}[p!]{0.467\linewidth}
\centering
\includegraphics[width=2.1in]{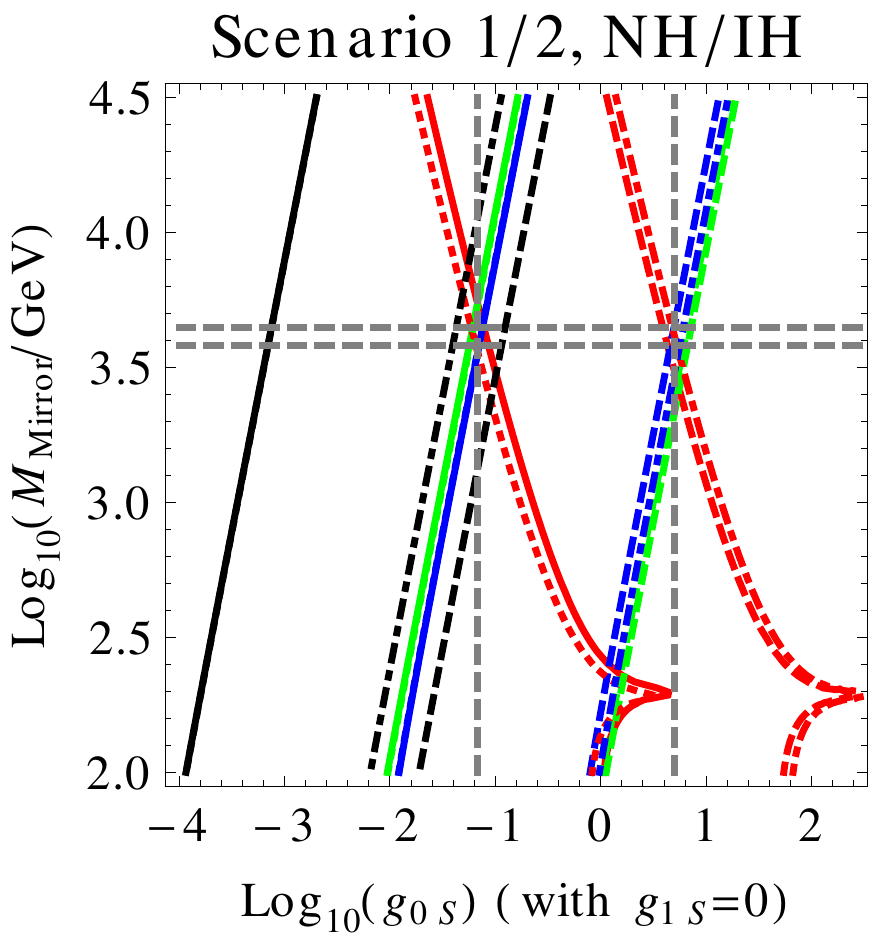}
\caption*{(a)}
\end{minipage}%
\begin{minipage}[p!]{0.467\linewidth}
\centering
\includegraphics[width=2.2in]{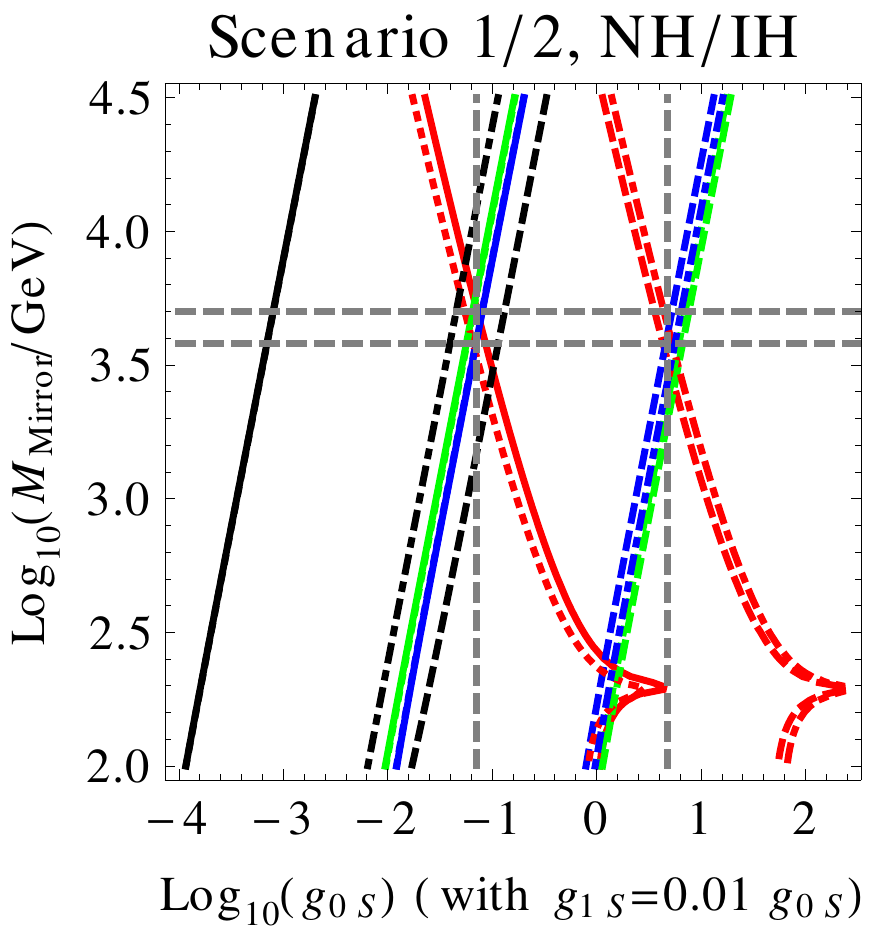}
\label{F1b}
\caption*{(b)}
\end{minipage}
\begin{minipage}[p!]{0.467\linewidth}
\centering
\includegraphics[width=2.2in]{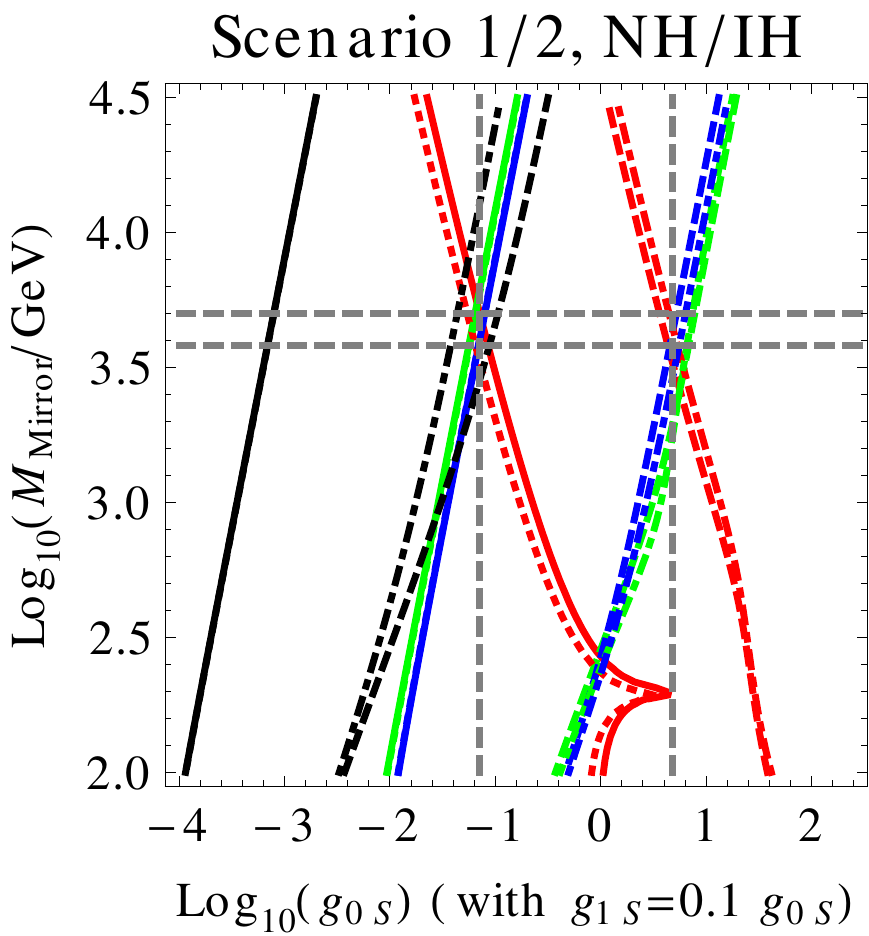}
\label{F1c}
\caption*{(c)}
\end{minipage}%
\begin{minipage}[p!]{0.467\linewidth}
\centering
\includegraphics[width=2.2in]{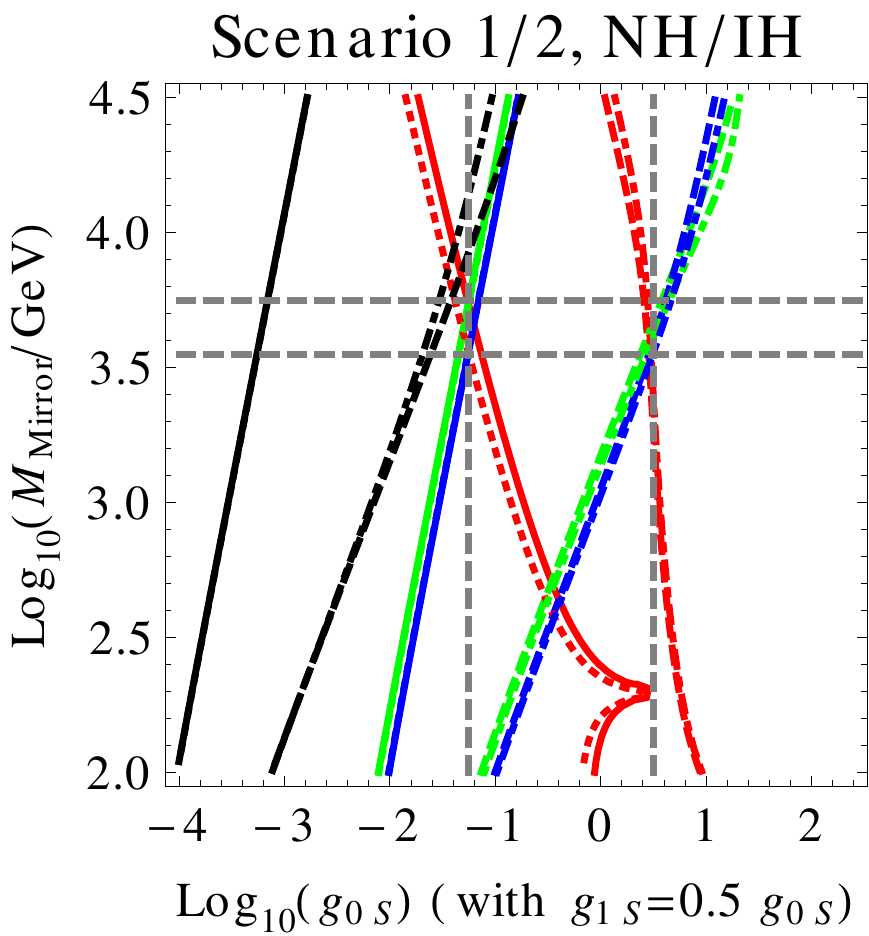}
\label{F1d}
\caption*{(d)}
\end{minipage}
\begin{minipage}[p!]{0.467\linewidth}
\centering
\includegraphics[width=2.1in]{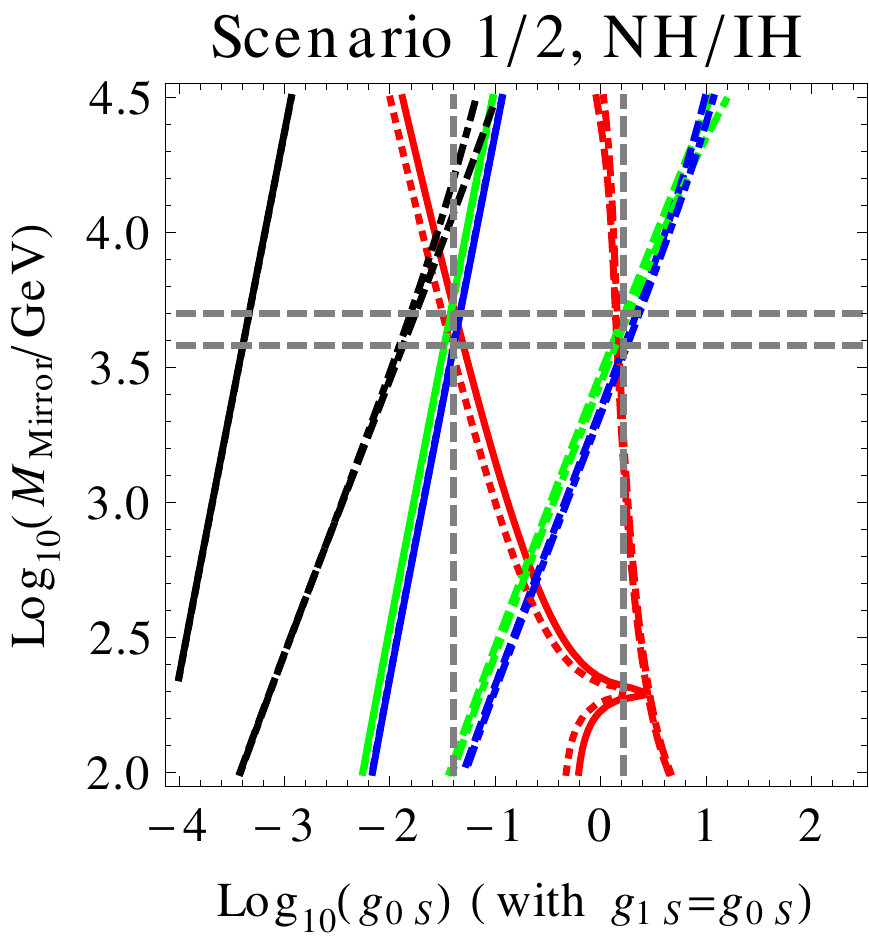}
\label{F1e}
\caption*{(e)}
\end{minipage}%
\begin{minipage}[p!]{0.467\linewidth}
\centering
\includegraphics[width=2.1in]{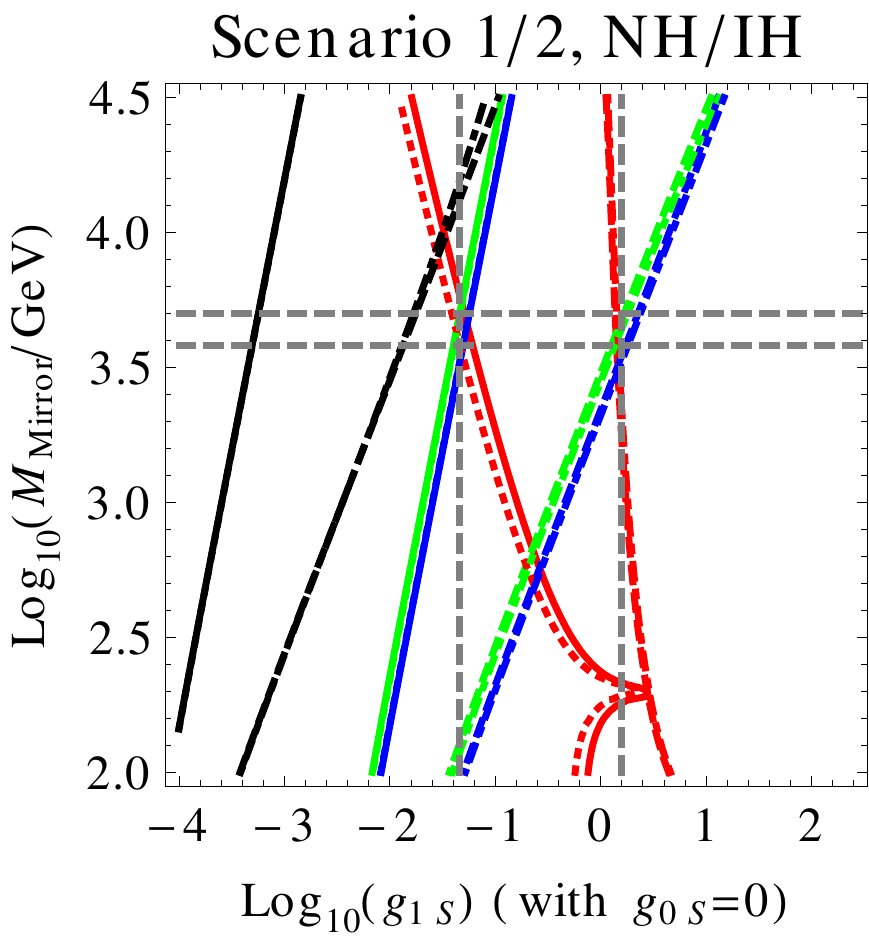}
\label{F1f}
\caption*{(f)}
\end{minipage}
\caption{\small
Contour plots of ${\cal B}(h \to \tau \mu) = 0.84 \%$ (red), 
${\cal B}(\mu \to e \gamma) = 5.7 \times 10^{-13}$ (black), 
${\cal B}(\tau \to \mu \gamma) = 4.4 \times 10^{-8}$ (blue) 
and ${\cal B}(\tau \to e \gamma) = 3.3 \times 10^{-8}$ (green) on the 
$({\rm Log}_{10}(M_{\rm mirror}/{\rm GeV}), {\rm Log}_{10}(g_{0S\, {\rm or} \, 1S}))$  plane for the {\it Dr. Jekyll} scenario. 
Solid: NH, S1; Dotted: IH, S1;
Dashed: NH, S2; Dot-dashed: IH, S2. 
See text in Sec. IV for details.}
\label{FigDrJekyll}
\end{figure}

In Fig.~(\ref{FigDrJekyll}), we plot the  contours of the branching ratios 
${\cal B} (h \to \tau \mu) =0.84 \%$ (red), 
${\cal B}(\mu \to e \gamma)=5.7 \times 10^{-13}$ (black),
${\cal B}(\tau\to \mu \gamma)=4.4 \times 10^{-8}$ (blue) and 
${\cal B}(\tau \to e \gamma) = 3.3 \times 10^{-8}$ (green)
on the $({\rm Log}_{10}(M_{\rm mirror}), {\rm Log}_{10}(g_{0S\, {\rm or} \, 1S}))$  plane 
for both Scenarios 1 and 2,  normal and inverted mass hierarchies and the 6 different 
cases of the Yukawa couplings (Figs.(\ref{FigDrJekyll}a)-(\ref{FigDrJekyll}f)) in the {\it Dr. Jekyll} scenario as specified by
Eqs.~(\ref{DrJekyllO})-(\ref{DrJekyllHiggses}). 
For the four lines with the same color (hence same process), 
solid and dashed lines are for Scenario 1 and 2 with normal mass hierarchy (NH) respectively,
while dotted and dot-dashed lines are for Scenario 1 and 2 with  inverted mass hierarchy (IH) respectively.

\begin{figure}[h!]
\begin{minipage}[p!]{0.467\linewidth}
\centering
\includegraphics[width=2.1in]{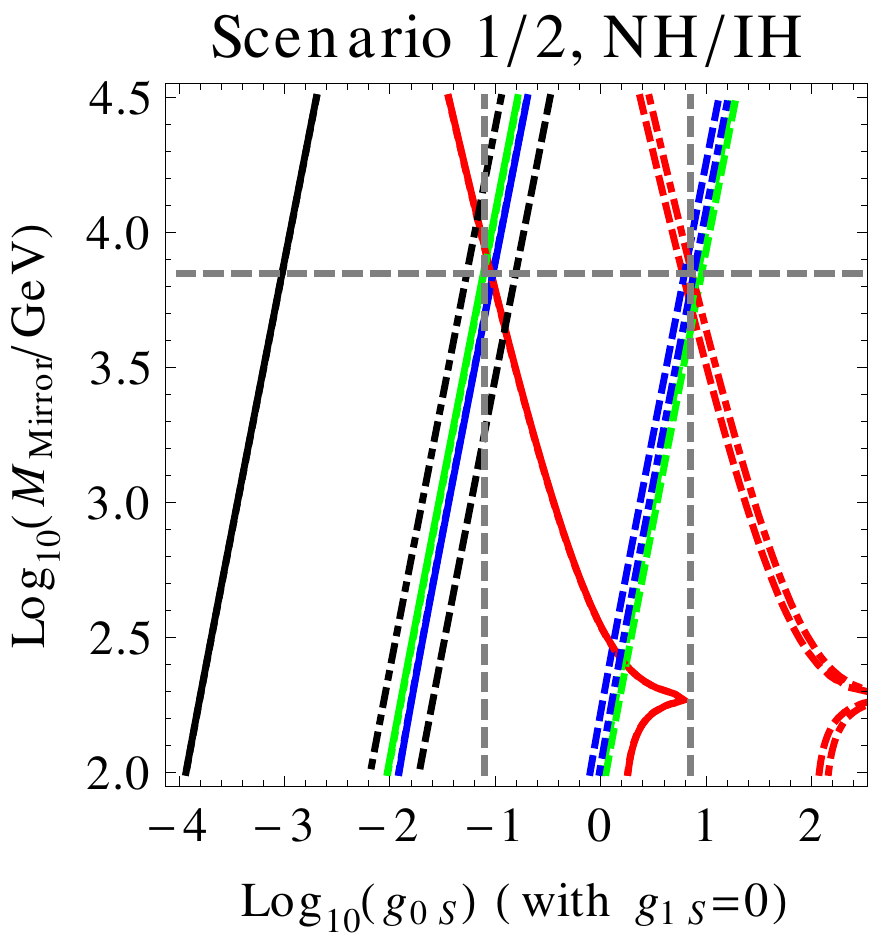}
\caption*{(a)}
\end{minipage}%
\begin{minipage}[p!]{0.467\linewidth}
\centering
\includegraphics[width=2.2in]{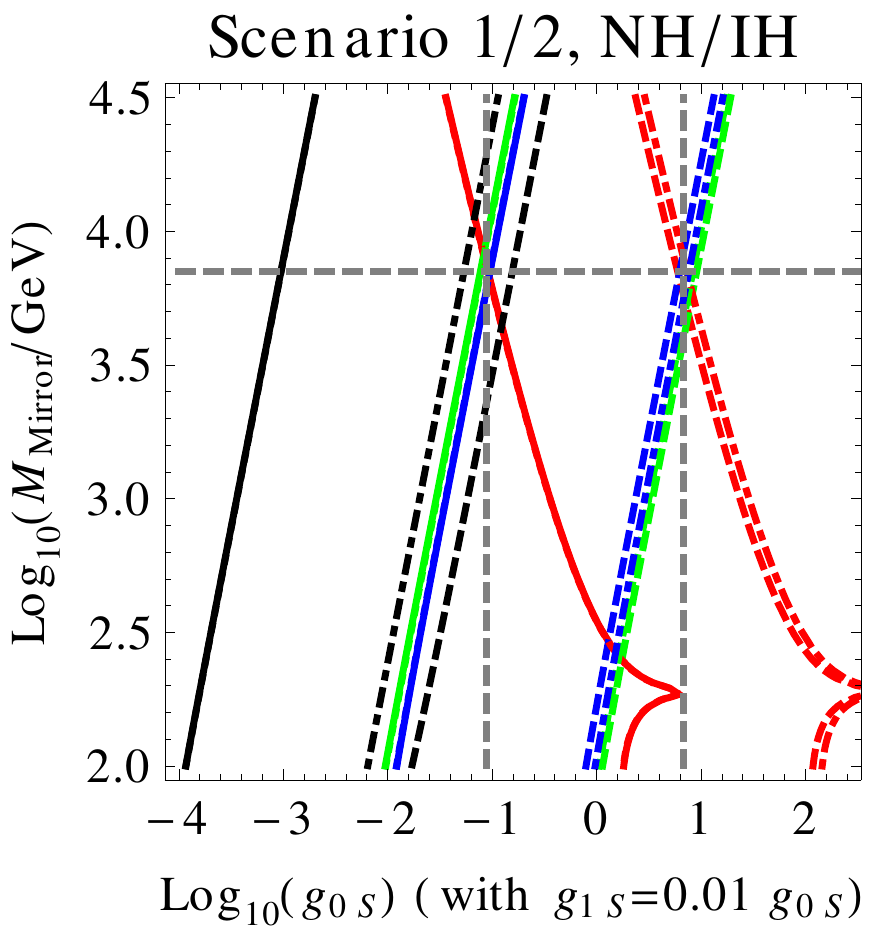}
\caption*{(b)}
\end{minipage}
\begin{minipage}[p!]{0.467\linewidth}
\centering
\includegraphics[width=2.2in]{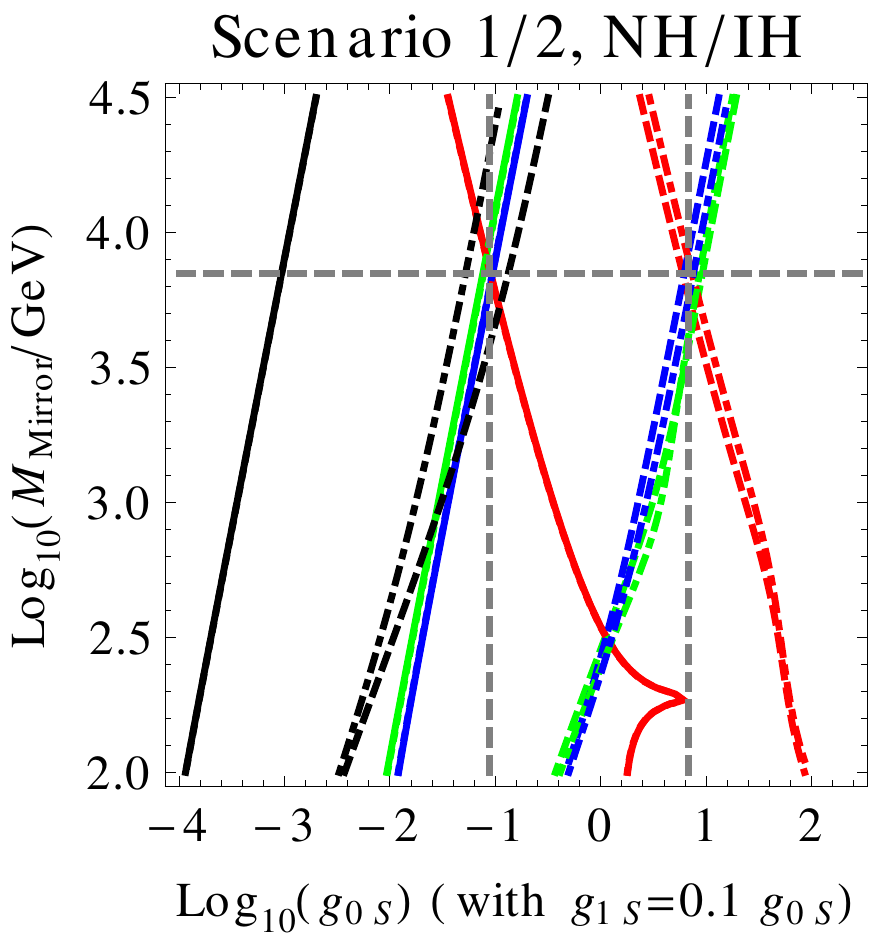}
\caption*{(c)}
\end{minipage}%
\begin{minipage}[p!]{0.467\linewidth}
\centering
\includegraphics[width=2.2in]{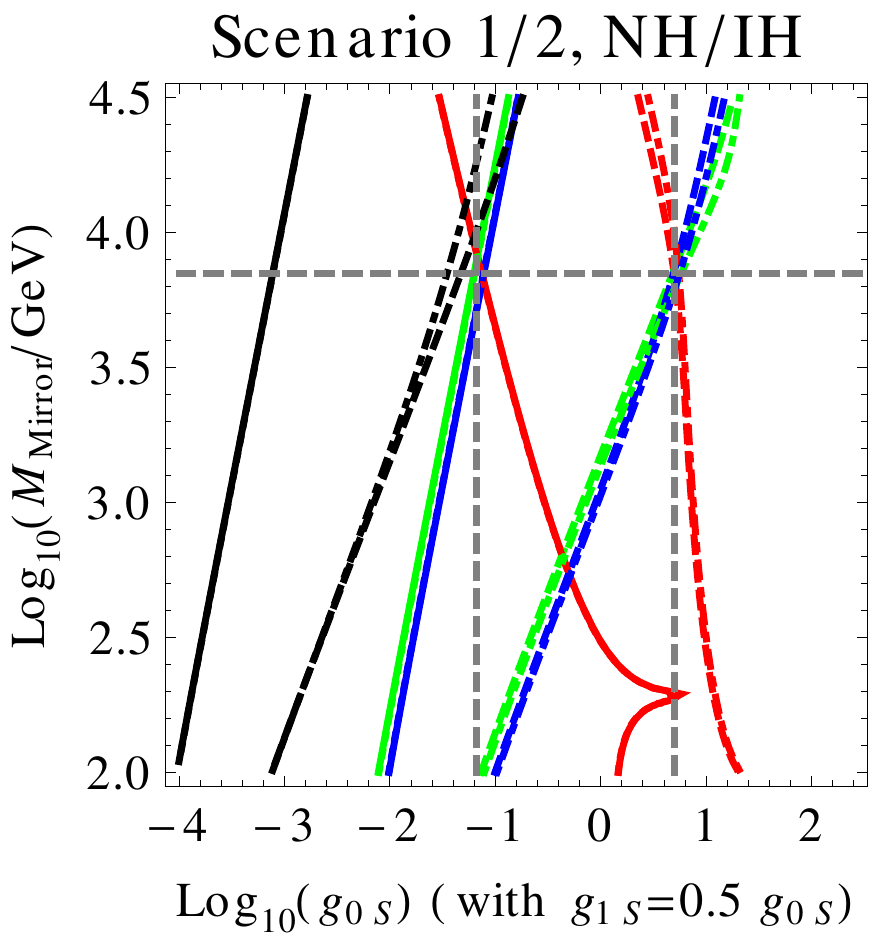}
\caption*{(d)}
\end{minipage}
\begin{minipage}[p!]{0.467\linewidth}
\centering
\includegraphics[width=2.1in]{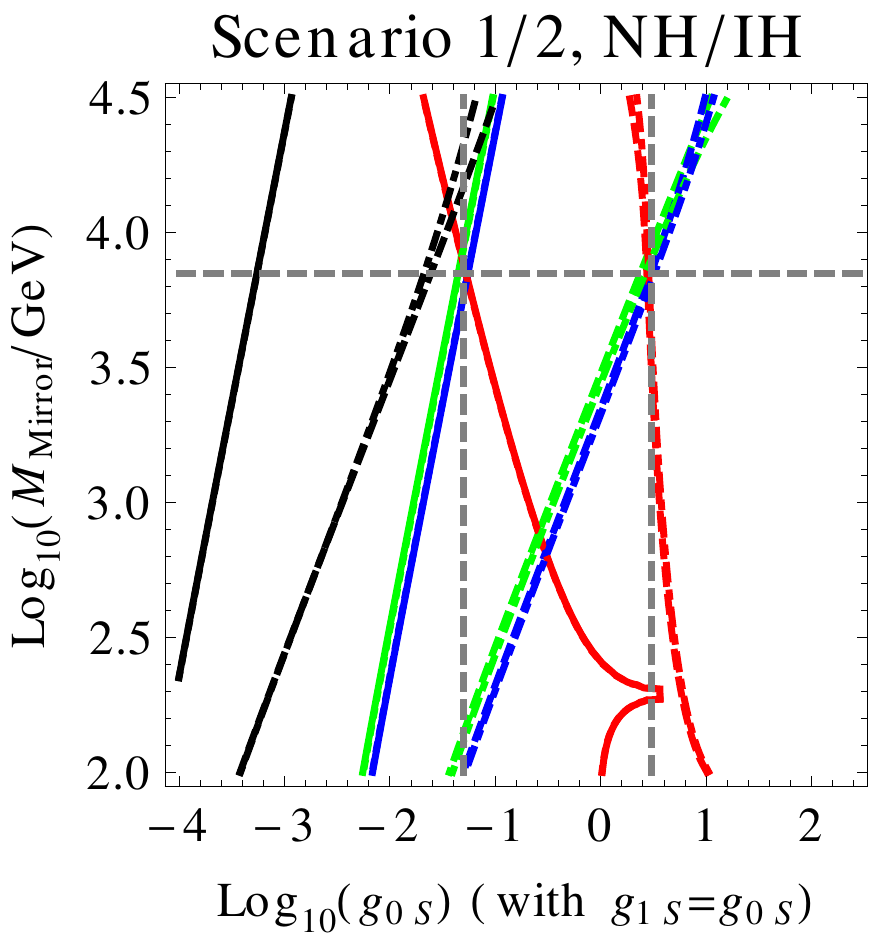}
\caption*{(e)}
\end{minipage}%
\begin{minipage}[p!]{0.467\linewidth}
\centering
\includegraphics[width=2.1in]{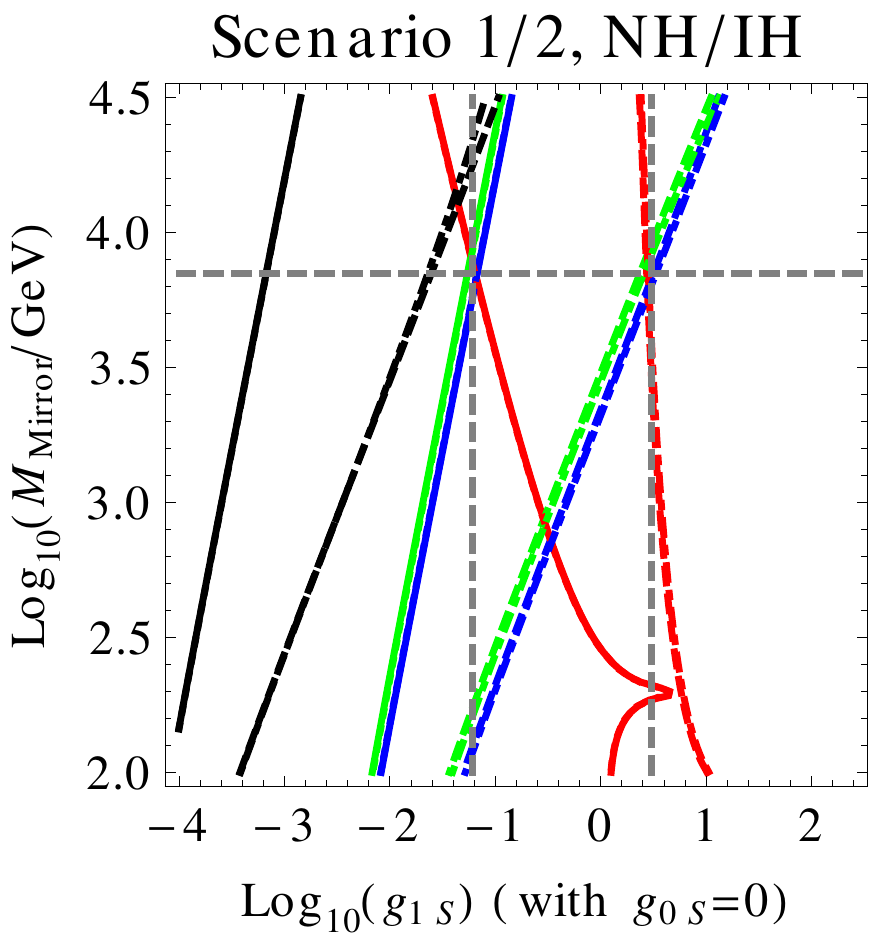}
\caption*{(f)}
\end{minipage}
\caption{
\small
Contour plots of ${\cal B}(h \to \tau \mu) = 0.84 \%$ (red), 
${\cal B}(\mu \to e \gamma) = 5.7 \times 10^{-13}$ (black), 
${\cal B}(\tau \to \mu \gamma) = 4.4 \times 10^{-8}$ (blue) 
and ${\cal B}(\tau \to e \gamma) = 3.3 \times 10^{-8}$ (green) on the 
$({\rm Log}_{10}(M_{\rm mirror}/{\rm GeV}), {\rm Log}_{10}(g_{0S\, {\rm or} \, 1S}))$  plane for the {\it Mr. Hyde} scenario. Solid: NH, S1; Dotted: IH, S1;
Dashed: NH, S2; Dot-dashed: IH, S2. 
See text in Sec. IV for details.}
\label{FigMrHyde}
\end{figure}

Figs.~(\ref{FigMrHyde}a)-(\ref{FigMrHyde}f) are the same as 
Figs.~(\ref{FigDrJekyll}a)-(\ref{FigDrJekyll}f)) respectively but for {\it Mr. Hyde} scenario as specified by
Eqs.~(\ref{MrHydeO})-(\ref{MrHydeHiggses}). 

By studying in details of all the plots in these two figures, we can deduce the following results:

\begin{itemize}

\item 
The bumps at $M_{\rm mirror} \sim $ 200 GeV at all the plots in these two figures 
are due to large cancellation in the amplitudes between the two one-particle reducible  
(wave function renormalization) diagrams and the irreducible one-loop diagram shown in Fig.~(\ref{FeynDiag}). 
As a result, the Yukawa couplings have to be considerable larger in the contour lines of fixed branching ratios 
of the processes.

\item 
For the two processes $\tau \to \mu \gamma$ (blue lines) and $\tau \to e \gamma$ 
(green lines) in all these plots, the solid and dotted lines are coincide to each other while 
the dashed and dot-dashed lines are very close together. Thus there are essentially no differences 
between the normal and inverted mass hierarchies in both Scenarios 1 and 2 in these two processes. 
However, for the process $\mu \to e \gamma$ (black lines), only the solid and dotted lines are coincide to each other.
Thus there are some differences between normal and inverted mass hierarchies in Scenario 2 but not in Scenario 1
for this process, in particular for cases (a)-(d) in which $g_{1S} \leq 0.5 g_{0S}$.

\item
For $h \to \tau \mu$ (red lines), the solid (dashed) and dotted (dot-dashed) lines are 
either very close (in Fig.~(\ref{FigDrJekyll}) 
for {\it Dr. Jekyll} scenario) or mostly coincide (in Fig.~(\ref{FigMrHyde}) for {\it Mr. Hyde} scenario).

\item
Note that the regions to the right side of the black, blue and green lines in all the plots in 
these two figures are excluded by the low energy limits of ${\cal B}(\mu \to e \gamma)$,
${\cal B}(\tau \to \mu \gamma)$ and ${\cal B}(\tau \to e \gamma)$ respectively.
The CMS result of ${\cal B}(h \to \tau \mu) = 0.84 \%$ (red lines), if not due to statistical fluctuations, 
is compatible with these low energy limits only if there are intersection points of the red lines 
with the corresponding black, blue and green lines.

Take Fig.~(2a) as an example. For case of {\it Dr. Jekyll} and in Scenario 1,
the solid (or dotted) red line intersects with the solid (or dotted) blue and green lines 
at $M_{\rm mirror} \sim $ 4.47 TeV  where $g_{0S} \sim $ 0.0676. 
In Scenario 2, the dashed (or dot-dashed) 
red line intersects the dashed (or dot-dashed) blue or green lines
at $M_{\rm mirror} \sim $ 3.55 TeV with a considerable larger $g_{0S} \sim$ 5.01.
For the black lines from the most stringent limit of $\mu \to e \gamma$, their intersections with the red lines 
are well beyond 10 TeV for the mirror lepton masses. 
Similar statements can be obtained from the other plots in these two figures. 
From these intersections in these figures, one can deduce the lower (upper) limits of the mirror 
fermion masses (couplings) which we summarize in Table~\ref{tab1}.
Such a large mirror lepton mass $M_{\rm mirror}$ or coupling $g_{0S}$ indicates a break down of the 
perturbative calculation and/or violation of unitarity. However taking what we have literally 
there is tension between the large branching ratio ${\cal B}(h \to \tau \mu)$ from LHC and the low energy limits 
of ${\cal B}(\tau \to (\mu,e) \gamma)$ and ${\cal B}(\mu \to e \gamma)$, in particular the latter one.

\item
In the event that the CMS result in Eq.~(\ref{LHCResults}) 
is just a statistical fluctuation, the limits in Eq.~(\ref{LHCLimits}) will be improved 
further in LHC Run 2. The contour lines of these future limits would be located 
to the left side of the current red lines in the two Figs.~(\ref{FigDrJekyll}) and (\ref{FigMrHyde}).
Their intersections with the black, blue and green lines would then be at lower mirror 
lepton masses and smaller Yukawa couplings, since the low energy limits of the LFV decays 
$l_i \to l_j \gamma$  are unlikely to be changed significantly anytime soon. Certainly this would alleviate
the tension mentioned above.

\end{itemize}
%
\setlength{\arrayrulewidth}{0.2pt}
\begin{table}[t]
\caption{The lower (upper) limit of mirror fermion masses (couplings).} 
\centering
\tabcolsep=8pt
\renewcommand{\arraystretch}{0.6}
\begin{tabular}{l c  c c c c }
&&\multicolumn{2}{c}{Scenario 1}& \multicolumn{2}{c}{Scenario 2}  \\
Mode & Quantity & {\it Dr.Jekyll} & {\it Mr. Hyde} & {\it Dr.Jekyll} & {\it Mr. Hyde} 
\\ [0.1ex]
\hline\hline
&Mass (TeV) & 4.47 & 7.08 & 3.55 & 7.08    \\[0.1ex]
\raisebox{2ex}{$\tau \to (\mu , e) \gamma$}
&$g_{0S} (g_{1S})$ & 0.07 & 0.09 & 5.01 & 6.76 \\[0.1ex]
\cline{1-6}
&Mass (TeV) & $\sim 100$ & $>10^{2.5}$ & $\sim 95$ & $>10^{2.5}$ \\[0.1ex]
\raisebox{2ex}{$\mu \to e \gamma$}
&$g_{0S} (g_{1S})$ & $10^{-2.6}$ & $10^{-2.1}$ & 0.16 & 0.40  \\[0.1ex]
\cline{1-6}
\hline
\end{tabular}
\label{tab1}
\end{table}
%

\section{Conclusion}

To summarize, CMS has reported excess in the charged lepton flavor violating 
Higgs decay $h \to \tau \mu$ at 2.4$\sigma$ level. 
More data is needed to collect at Run 2 so as to confirm whether these are indeed true signals 
or simply statistical fluctuations.

If the branching ratio of $h \to \tau \mu$ is indeed at the percent level, new physics associated 
with lepton flavor violation may be at a scale not too far from the electroweak scale. Crucial question 
is whether this large branching ratio of $h \to \tau \mu$ is compatible with 
the current low energy limits of $\tau \to \mu \gamma$ and $\tau \to e \gamma$ from Belle experiments 
and the most stringent limit of $\mu \to \gamma$ from MEG experiment. 

We analyze these lepton flavor violating processes in the context of an extended mirror fermion model 
with non-sterile electroweak scale right-handed neutrinos as well as a horizontal $A_4$ symmetry imposed
on the lepton sector. We found that the masses of the mirror lepton fermions entering the loops of these processes
can be of the order of a few hundred GeV to a few TeV depending on the sizes of the 
Yukawa couplings among the leptons, mirror leptons and the scalar singlets in the model 
as well as whether or not the 125 GeV scalar boson is a Higgs impostor and 
which scenario one assumes for the three unknown PMNS-type mixing matrices. 
We demonstrate that in general there is tension between the LHC result and the low energy limits since 
these results are compatible only if the mirror lepton masses are quite heavy and/or the Yukawa couplings 
involving the scalar singlets are large. 

Before we depart, we comment on the possible collider signals for the mirror fermions~\cite{Chakdar:2015sra}. 
Mirror leptons if not too heavy can be produced at the LHC via electroweak 
processes~\cite{Hung:2006ap}, {\it e.g.} 
$q \bar q \to Z \to l_R^M \overline{ l_R^M}, \nu_R \overline{\nu_R}$ and 
$q \overline{q^\prime} \to W^\mp \to l^M_{R} \overline{\nu_{R}}, \nu_{R}\overline{l^M_{R}}$.
The mirror lepton decays as $l_R^M \to l_L + \phi_{S}$ or $l^M_R \to  \nu_R + W^{-(*)}$ 
for $m_{l^M_R} > m_{\nu_R}$ plus the conjugate processes, 
while the right-handed neutrino can decay as $\nu_R \to \nu_L + \phi_S$ or 
$\nu_R \to l^M_R + W^{+(*)}$ for $m_{\nu_R} > m_{l^M_R}$ followed by $l_R^M \to l_L + \phi_{S}$.
If kinematics allowed, the scalar singlet $\phi_{S}$ can decay into lepton pair as well through mixings; 
otherwise they would appear as missing energies like neutrinos.
Thus the signals at the LHC or future 100 TeV SPPC would be multiple lepton pairs plus missing energies. 
In the case where the right-handed neutrinos are Majorana fermions, 
we would have same sign dilepton plus missing energies.
Assuming $l_R^M \to l_L + \phi_{S}$ is the dominant mode and the Yukawa couplings are small enough,
the decay length of the mirror lepton could be as large as a few millimeter~\cite{Chakdar:2015sra}.  
Thus the mirror lepton may lead to a displaced vertex and decay outside the beam pipe.
These leptonic final states may have been discarded by the current algorithms adopted 
by the LHC experiments.  
It is therefore quite important for the experimentalists to devise new algorithms to search for 
these mirror fermions that may decay outside the beam pipe.

The scale of new physics may be hidden in the lepton flavor violating processes 
like $h \to \tau (\mu,e)$, $\tau \to (\mu,e) \gamma$, $\mu \to e \gamma$, $\mu \to eee$,
$\mu$-$e$ conversion {\it etc.} 
Ongoing and future experiments at high energy and high intensity frontiers
could shed light in the mirror fermion model that may responsible to these
lepton flavor violating processes.

\section*{Acknowledgments}
We would like to thank P. Q. Hung for useful discussions.
This work was supported in part by the Ministry of Science and Technology (MoST) of Taiwan under
grant numbers 104-2112-M-001-001-MY3.


\section*{Appendix}

The dimensionless coefficients $C^{aij}_L$ and $C^{aij}_R$ defined in Eq.~(\ref{AmpAB}) are given by
\begin{eqnarray}
C^{aij}_L & = & \frac{g O_{a1}}{2 s_2 m_W (m_i^2 - m_j^2)}  \sum_{k,m} \int_0^1 d x
\biggl\{ \left[
(1 - x) \left( m_i m_j^2 \mathcal{U}^{L \, k}_{im} \left( {\mathcal U}^{L \, k}_{mj} \right)^*
+ m_j m_i^2 \mathcal{U}^{R \, k}_{im} \left( {\mathcal U}^{R \, k}_{mj} \right)^* \right) \nonumber \right. \biggr. \\
& + & \left. \left.  m_i m_j M_m \mathcal{U}^{L \, k}_{im} \left( {\mathcal U}^{R \, k}_{mj} \right)^* \right] 
\log \left( \frac{\Delta_1}{\Delta_2} \right) 
+ M_m \mathcal{U}^{R \, k}_{im} \left( {\mathcal U}^{L \, k}_{mj} \right)^* 
\left( m_i^2 \log \Delta_1 - m_j^2 \log \Delta_2 \right)
\right\} \nonumber \\
& + & \frac{g O_{a2}}{2 s_{2M} m_W} \sum_{k,m}   M_m \mathcal{U}^{R \, k}_{im} \left( {\mathcal U}^{L \, k}_{mj} \right)^* 
\left(  - \frac{1}{2} - 2 \int_0^1 dx  \int_0^{1-x} dy \log \Delta_3 \right) \nonumber \\
& - &  \frac{g O_{a2} }{2 s_{2M} m_W} \sum_{k,m} M_m \int_0^1 dx  \int_0^{1-x} dy \frac{1}{\Delta_3}
\left\{ (1- 2y) \frac{m_i M_m}{m^2_{\widetilde H_a}} 
\mathcal{U}^{L \, k}_{im} \left( {\mathcal U}^{L \, k}_{mj} \right)^*  \right. \nonumber \\
&&  \, + \, (1- 2x) \frac{m_j M_m}{m^2_{\widetilde H_a}} \mathcal{U}^{R \, k}_{im} \left( {\mathcal U}^{R \, k}_{mj} \right)^*
+ (1-x-y) \frac{m_i m_j}{m^2_{\widetilde H_a}} \mathcal{U}^{L \, k}_{im} \left( {\mathcal U}^{R \, k}_{mj} \right)^* \nonumber \\
&& \Biggl. - \left[ x y + (1-x-y) (y  r_i + x  r_j) - r_m \right] 
\mathcal{U}^{R \, k}_{im} \left( {\mathcal U}^{L \, k}_{mj} \right)^* 
\Biggr\} \; ,
\end{eqnarray}
$C^{aij}_R$ can be obtained from $C^{aij}_L$ simply by substituting 
$\mathcal{U}^{L} \leftrightarrow \mathcal{U}^{R}$, namely
\begin{eqnarray}
C^{aij}_R & = & \frac{g O_{a1}}{2 s_2 m_W (m_i^2 - m_j^2)}   \sum_{k,m} \int_0^1 d x
\biggl\{ \left[
(1 - x) \left( m_i m_j^2 \mathcal{U}^{R \, k}_{im} \left( {\mathcal U}^{R \, k}_{mj} \right)^*
+ m_j m_i^2 \mathcal{U}^{L \, k}_{im} \left( {\mathcal U}^{L \, k}_{mj} \right)^* \right) \nonumber \right. \biggr. \\
& + & \left. \left.  m_i m_j M_m \mathcal{U}^{R \, k}_{im} \left( {\mathcal U}^{L \, k}_{mj} \right)^* \right] 
\log \left( \frac{\Delta_1}{\Delta_2} \right) 
+ M_m \mathcal{U}^{L \, k}_{im} \left( {\mathcal U}^{R \, k}_{mj} \right)^* 
\left( m_i^2 \log \Delta_1 - m_j^2 \log \Delta_2 \right)
\right\} \nonumber \\
& + & \frac{g O_{a2}}{2 s_{2M} m_W} \sum_{k,m}  M_m \mathcal{U}^{L \, k}_{im} \left( {\mathcal U}^{R \, k}_{mj} \right)^* 
\left(  - \frac{1}{2} - 2 \int_0^1 dx  \int_0^{1-x} dy \log \Delta_3 \right) \nonumber \\
& - &  \frac{g O_{a2}}{2 s_{2M} m_W}  \sum_{k,m} M_m \int_0^1 dx  \int_0^{1-x} dy \frac{1}{\Delta_3}
\left\{ (1- 2y) \frac{m_i M_m}{m^2_{\widetilde H_a}} \mathcal{U}^{R \, k}_{im} \left( {\mathcal U}^{R \, k}_{mj} \right)^*  \right. \nonumber \\
&&  \, + \, (1- 2x) \frac{m_j M_m}{m^2_{\widetilde H_a}} \mathcal{U}^{L \, k}_{im} \left( {\mathcal U}^{L \, k}_{mj} \right)^*
+ (1-x-y) \frac{m_i m_j}{m^2_{\widetilde H_a}} \mathcal{U}^{R \, k}_{im} \left( {\mathcal U}^{L \, k}_{mj} \right)^* \nonumber \\
&& \Biggl. - \left[ x y  + (1-x-y) (y  r_i + x  r_j) - r_m \right] 
\mathcal{U}^{L \, k}_{im} \left( {\mathcal U}^{R \, k}_{mj} \right)^* 
\Biggr\} \; .
\end{eqnarray}
The $\Delta_1$, $\Delta_2$ and $\Delta_3$ are given by
\begin{eqnarray}
\Delta_1 & = & x r_m + (1-x) r_k - x (1-x) r_j - i 0^+ \; , \\
\Delta_2 & = & x r_m + (1-x) r_k - x (1-x) r_i - i 0^+ \; , \\
\Delta_3 & = & (x + y) r_m + (1-x-y) (r_k - y r_i - x r_j ) - x y  - i 0^+ \; .
\end{eqnarray}
Here $r_m = M_m^2/m_{\widetilde H_a}^2$, 
$r_{i,j} = m_{i,j}^2/m_{\widetilde H_a}^2$ and $r_k = m_k^2/m_{\widetilde H_a}^2$  
with $M_m$, $m_{i,j}$ and $m_k$ denoting the masses of the mirror leptons, leptons and scalar singlets respectively.


\end{document}